\documentstyle[openbib,11pt,psfig]{article}

\newcommand{\eq}{\begin{equation}}
\newcommand{\en}{\end{equation}}
\newcommand{\la}{\label}
\newcommand{\p}{\partial}
\begin{document}

\noindent {\bf An Idealized Pulsar Magnetosphere: the 
              Relativistic Force-Free Approximation}\\

\noindent Simon P. Goodwin$^{1,2}$, Jonathan Mestel$^{3}$, Leon
Mestel$^{1}$, Geoffrey
A.E. Wright$^{1}$\\
{\em $^{1}$Astronomy Centre, Division of Physics and Astronomy,
University of Sussex, Falmer, Brighton BN1 9QJ}pine\\
{\em $^{2}$Department of Physics and Astronomy, Cardiff University, 5
The
Parade, Cardiff} CF24 3YB\\
{\em $^{3}$Department of Mathematics, Imperial College of Science,
Technology and Medicine,
      Prince Consort Rd., London SW7 2BZ}\\

Received

{\bf Abstract} The non-dissipative relativistic force-free
condition should be a good approximation to describe the
electromagnetic field in much of the pulsar magnetosphere, but we
may plausibly expect it to break down in singular domains.
Self-consistent magnetospheric solutions are found with field
lines closing both at and within the light-cylinder. In general,
the detailed properties of the solutions may be affected
critically by the physics determining the appropriate choice of
equatorial boundary condition beyond the light-cylinder.

\section{Introduction}

   With canonical values inserted for the neutron star dipole moment and
the particle density, and allowing for realistic pair production
efficiency near the star in the wind zone, the electromagnetic
energy density in the pulsar magnetosphere will be much greater
than the kinetic energy density except for very high values of the
particle $\gamma$-values. This suggests that over much of the
magnetosphere, the {\em relativistic force-free equation} will be
a good approximation for determining the magnetic field structure.
Equally, experience with the analogous non-relativistic equation
suggests that there are likely to be local domains in which
non-electromagnetic forces (including possibly inertial forces)
are required for force balance. An associated question is whether
the dissipation-free conditions can be imposed everywhere, or
whether the dynamics of the problem will itself demand local
breakdown in the simple plasma condition ${\bf E}\cdot {\bf B} = 0$ in
singular domains, in addition to the acceleration/pair production
domain.

    Of the various discussions of the axisymmetric
pulsar magnetosphere in the literature, which go back to the early
days of pulsar theory (e.g. Scharlemann and Wagoner 1973), we
refer particularly to the recent paper by Contopoulos {\em et al.}
1999, from now on referred to as CKF, and to those by Michel
1973a,b, 1991, Mestel and Wang (MW) 1979, Mestel {\em et al.}
(MRWW) 1985, Fitzpatrick and Mestel (FM) 1988a,b , Mestel and
Pryce (MP) 1992, and Mestel and Shibata (MS) 1994. (Much of the
relevant material appears in Mestel 1999, 2003.) Most models have
within the light-cylinder (`l-c') a `dead zone,' with field lines
that close within the l-c, and without any current flow along the
(purely poloidal) field, and a `wind zone,' with poloidal field
lines that cross the l-c, and with poloidal currents maintaining a
toroidal field component. The spin-down of the star occurs through
the action of the magnetic torques associated with the circulatory
flow of current across the light-cylinder.

Although we are here discussing an idealised mathematical model of
the pulsar magnetosphere -- not even allowing the pulsar to pulse!
-- its structure may nevertheless have great relevance to the
class of real pulsars whose magnetic axis is not far from
alignment. A striking example of this is PSR0826-34 (Biggs et al
1985), a pulsar which emits throughout the entire pulse period,
suggesting our line of sight is almost along the rotational axis.
Other bright pulsars which are suspected on the basis of their
profile widths and relatively slow spin-down rates to be near
alignment include PSR0031-07, PSR0943+10 and PSR0809+74 (see
Lyne and Manchester 1988 and the catalogue in Rankin (1993)).

All these pulsars exhibit highly regular subpulse `drift`, by
which the sub-pulses march gradually and steadily through the
pulse window at a fixed rate. This suggests that their
magnetospheres are in a quasi-steady state, and that steady-state
models such as those discussed here may have direct physical
relevance. Furthermore, the fact that all the above pulsars
occasionally - yet apparently spontaneously - switch to one or
more alternate `drift` rates, accompanied by a widening (or
narrowing) of the pulsar emission profile (Wright and Fowler 1981,
Vivekanand and Joshi 1997, van Leeuwen et al 2002) would indicate 
that the new steady
emission has moved to outer (inner) field lines. Thus it would
appear that more than one steady state is possible in the same
pulsar - presumably corresponding to different polar cap shapes,
and hence to different `dead zone' boundaries. This would add
relevance and motivation to our discussion in Section 6, where we
derive magnetosphere solutions for a range of assumptions about
the location and properties of the dead zone boundary.

\section{The relativistic force-free equation}
   The outer crust of the neutron star of radius $R$ rotates with angular 
velocity $\alpha$ about the axis defined by the unit vector {\bf k}. We
confine attention in this paper to steady, axisymmetric states,
and with the magnetic axis parallel rather than anti-parallel to
the rotation axis ${\bf k}$. The notation is as in the cited
papers (e.g. FM). The cylindrical polar coordinate system
$(\varpi, \phi, z)$ is based on ${\bf k}$; ${\bf t}$ is the unit
azimuthal, toroidal vector.
  The poloidal magnetic field ${\bf B}_{{\rm p}}$ is everywhere described by 
the flux function $P(\varpi,z)$:
\begin{equation}
\label{1}
{\bf B}_{{\rm p}} = - \nabla P\times \left(\frac{{\bf t}}{\varpi}\right)
= \frac{1}{\varpi}\left(\frac{\partial P}{\partial z},\, 0,
\,-\frac{\partial P}{\partial \varpi}\right).
\end{equation}
By Amp\`ere's law, ${\bf B}_{{\rm p}}$ is maintained by the toroidal
current density
\begin{eqnarray}
\label{1.5}
\label{1.6}
  {\bf j}_{{\rm t}} & = &(c/4\pi)(\nabla \times {\bf B}_{{\rm p}}) =
(c/4\pi)(\nabla \times {\bf B})_{\phi}{\bf t}\\
                    & = &(c/4\pi)[\nabla^{2}P/\varpi - (2/\varpi^{2})
\p P/\p\varpi]{\bf t}.
\end{eqnarray}
The basic, unperturbed stellar field is assumed to have a dipolar angular 
dependence and polar field strength $B_s$ . If as assumed, 
$\alpha R/c \ll 1$, it will be found that the 
magnetospheric currents have little effect on the form of $P$ {\em near the 
star}, so that it is then well approximated locally by the vacuum form
\eq
\la{1.4}
P = -\frac{B_{s}R^{3}}{2}\frac{\varpi^{2}}{(\varpi^{2} + z^{2})^{3/2}}.
\en

   In the wind zone of an active magnetosphere there
will be also a toroidal component ${\bf B}_{{\rm t}}$, conveniently
written
\begin{equation}
\label{2}
B_{\phi}{\bf t} = -[4\pi S/c]({\bf t}/\varpi),
\end{equation}
where $S$ clearly must vanish on the axis. By Amp\`ere's law, the field
(\ref{2}) is maintained by the poloidal
current density
\begin{equation}
\label{3}
{\bf j}_{{\rm p}} =(c/4\pi)\nabla \times {\bf B}_{{\rm t}} = - \nabla S
\times {\bf t}/\varpi.
\end{equation}
Thus $S(\varpi,z)$ is a Stokes stream function, constant on the poloidal
current lines, with
$-2\pi S$  measuring the total outflow of charge between the axis and
the
current line $S$. In
a steady state the total outflow of charge must be zero,
so there must be a current closing streamline on which $S$ vanishes.

  The crust is taken to be a classical perfect conductor, with the
electric field
\begin{equation}
\label{4}
{\bf E} = - (\alpha\varpi/c) {\bf t} \times {\bf B}.
\end{equation}
In the surrounding magnetosphere, {\bf E} is in general written as
the sum of a `corotational' part of the form (\ref{4}) and a
`non-corotational' part $\nabla \psi$:
\eq \la{4.1}
{\bf E} = - (\alpha \varpi/c){\bf t} \times {\bf B} - \nabla \psi,
\en
where the function $\psi$ is determined by the magnetospheric
physics.

  A systematic treatment of the dynamics of a `cold,'
dissipation-free electron-positron gas has been given by Melatos
and Melrose (1996) (see also Blackman and Field 1993). The
significant differences from a normal plasma are due in part to
the particles being relativistic, and also through both species
having the same rest-mass, instead of differing by a factor
$Am_{{\rm H}}/m_{{\rm e}}$ in standard notation. The two-fluid,
collision-free and so non-dissipative equations, written in terms
of `lab-frame' number densities $n^{\pm}$ and  velocities ${\bf
v}^{\pm}$, and Lorentz factors $\gamma^{\pm}  =\{1 -
(v^{\pm}/c)^2\}^{-1/2}$, are reduced to one-fluid equations in
terms of
\begin{eqnarray}
\la{4.2}  n  = (n^+ + n^-),\quad &{\bf U}& = \frac{n^+{\bf v}^+
+ n^-{\bf v}^-}{n^+
  + n^-},\quad\quad \Gamma  = (1 - U^2/c^2)^{-1/2} \nonumber\\
  \rho_{{\rm e}} & =& e(n^+ - n^-),\quad\quad {\bf j}  = e(n^+{\bf v}^+
- n^-{\bf  v}^-)
\end{eqnarray}
(The lab-frame densities $n^{\pm}$ are related to the proper
densities $n_0^{\pm}$ by $n^{\pm} = \gamma^{\pm}n_0^{\pm}$.)

The particle continuity and current continuity equations retain
their standard forms \eq \la{4.3}
  \frac{\p n}{\p t} +{\bf\nabla}\cdot(n{\bf U}) =0;\quad\quad \frac{\p
\rho_{{\rm
e}}}{\p t} + {\bf \nabla \cdot j} = 0. \en
  The equation of
bulk motion with velocity ${\bf U}$ reduces to
\eq \la{4.4}
  mn\frac{\p(\Gamma {\bf U})}{\p t}+ m n{\bf U \cdot
\nabla}(\Gamma {\bf U}) =  \frac{({\bf j} - \rho_{{\rm e}}{\bf
U}){\bf \times B}}{c} \en
provided $j \ll neU$, even if, as
expected in a relativistic problem, the `convection current'
$\rho_{{\rm e}}U$ due to bulk motion of the net charge density is
not small compared with the current due to relative motion of the
two species. The generalized Ohm's law reduces to
\eq \la{4.5}
\frac{m}{e^2}\frac{\p}{\p t}\left(\frac{\Gamma {\bf j}}{n}\right)
+\frac{m}{e^2}{\bf U \cdot \nabla}\left(\frac{\Gamma{\bf
j}}{n}\right) + \frac{m}{ne^2}({\bf j}-\rho_{{\rm e}}{\bf U}){\bf \cdot
\nabla}(\Gamma {\bf U})  = {\bf E} + \frac{{\bf U \times
B}}{c}.\en

As stated, in this paper we are interested primarily in steady
states that are force-free over the bulk of the domain. Just as in
a classical plasma, in an $\pm e$-plasma the total particle number
density will normally greatly exceed the GJ number density,
defined by (\ref{5}) below, so the `perfect conductivity' and the
`force-free' conditions are distinct: a gas satisfying the
`perfect conductivity' condition need not be force-free. However,
one can easily check that if inertial terms on the left of
(\ref{4.4}) are small compared with either term on the right, then
the inertial corrections in (\ref{4.5}) are a fortiori small. Thus
in the absence of collisions, radiation damping or
electron-positron annihilation -- all neglected in (\ref{4.4}) and
(\ref{4.5}) -- the force-free condition implies also the ideal MHD
(perfect conductivity) condition. To sum up: for our problem, the
appropriate approximations to the dynamical equations are the
ideal MHD condition
\eq \la{4.6}
{\bf E} + {\bf U \times B}/c =
0, \en
implying ${\bf E}\cdot{\bf B} = 0$, and the mutual cancellation of
the electric and magnetic contributions to the Lorentz force, i.e.
the vanishing of the term on the right of (\ref{4.4}),
\eq \la{4.7}
  \frac{({\bf j} - \rho_{{\rm e}}{\bf U})\times {\bf B}}{c}
  = \rho_{{\rm e}}{\bf E} + \frac{{\bf j \times B}}{c} = 0,
\en
on use of (\ref{4.6}).

   It should be noted that in the more general, non-axisymmetric case,
with {\bf k} and {\bf p} non-aligned, the time-dependent terms
will not be negligible everywhere. As discussed by Melatos and
Melrose (1996), at a finite distance far beyond the l-c, the
displacement current will dominate over the particle current, and
the MHD approximations must break down.

We limit discussion explicitly to the case in which the simple
plasma condition ${\bf E \cdot B} =  0$ is supposed to hold
throughout the magnetosphere, all the way from the rigidly
rotating, perfectly conducting star out to the l-c (the `inner
domain'), and beyond into the `outer domain' between the l-c and
infinity. Then from (\ref{4.1}), ${\bf B}\cdot\nabla \psi = 0$,
and the constant value of $\psi$ throughout the stellar crust,
implied by (\ref{4}), is propagated into the magnetosphere,
yielding
\begin{equation}
\la{4.8} {\bf E} = - (\alpha\varpi/c) {\bf t \times B}=
(\alpha \varpi/c)(-B_{z}, 0, B_{x}) = (\alpha/c){\bf \nabla} P, \en
so that all field lines corotate with the star. This assumes that within
the {\em dead} zone, there is no vacuum gap separating the negatively
and
positively charged domains (Holloway and Pryce 1981; MRWW; FM).
  In the open field line domain, near the star there has to be a
locally non-trivial component of ${\bf E}$ along ${\bf B}_{{\rm
p}}$, able to accelerate the primary electrons to $\gamma$-values
high enough for pair production to occur. If, as assumed in this
paper, the electron-positron plasma does achieve a steady state
(cf. Shibata et al 1998, 2002), the electric field will again
satisfy ${\bf E \cdot B}= 0$ but will now be given by ${\bf E} = -
\tilde{\alpha}(P)(\varpi/c){\bf t \times B}$ -- i.e. with the
field lines beyond the acceleration domain having individual
rotation rates $\tilde{\alpha}(P)$ that differ somewhat from the
rotation $\alpha$ of the star (cf. MS, Section 4). However, at
least for the more rapid rotators this effect will be fairly
small, and will for the moment be ignored.

  By the Poisson-Maxwell equation, the Goldreich-Julian  (1969)  (GJ)
charge density maintaining the electric field (\ref{4}) is
\begin{eqnarray}
\label{5} \rho_{{\rm e}} = \frac{\nabla\cdot{\bf E}}{4\pi} &=&
-\frac{\alpha}{2\pi c}{\bf k}\cdot\left[{\bf B} - \frac{1}{2}{\bf
r}\times(\nabla \times {\bf B})\right]
\nonumber \\
&=& - \frac{\alpha}{2\pi c}\left[B_{z} - \frac{1}{2}\varpi(\nabla
\times {\bf B})_{\phi}\right].
\end{eqnarray}
With the rotation and magnetic axes aligned, the primary outflowing
particles are the negatively charged electrons, yielding a negative
current ${\bf j}_{\rm p}$ emanating from the polar cap. The stream
function $S$ defined by (\ref{3}) begins by increasing from zero on the
axis,
so that  $B_{\phi}$ is negative
-- the field lines are twisted backwards with respect to the axis ${\bf
k}$.
As the electric force density
\begin{equation}
  \label{5.5}
\rho_{{\rm e}}{\bf E} = - \rho_{{\rm e}}(\alpha \varpi/c){\bf t}\times
{\bf B}_{{\rm p}}
\end{equation}
is purely poloidal, in a force-free magnetosphere
the toroidal component of the magnetic force density ${\bf j}_{{\rm
p}}\times {\bf B}_{{\rm p}}/c$
vanishes (the `torque-free' condition), so ${\bf j}_{{\rm p}}$ must be
parallel to ${\bf B}_{{\rm p}}$,
yielding from (\ref{1}) and (\ref{3}) the functional relation
\begin{equation}
\label{6}
                   S = S(P), \quad {\bf j}_{{\rm p}}= \frac{dS}{dP}{\bf
B}_{{\rm p}}:
\end{equation}
the poloidal current streamlines are identical with the poloidal
field lines. The respective contributions of ${\bf B}_{{\rm p}}$
and ${\bf B}_{{\rm t}}$ to the poloidal force density are
\begin{equation}
\label{7}
{\bf j}_{{\rm t}}\times {\bf B}_{{\rm p}}/c =
(\nabla \times{\bf B})_{\phi}({\bf t}\times {\bf B}_{{\rm p}})/4\pi
\end{equation}
and
\eq
\label{8}
{\bf j}_{{\rm p}}\times {\bf B}_{{\rm t}}/c = \frac{4\pi
S}{c^{2}\varpi}\frac{{\rm d}S}{{\rm d}P}({\bf t}\times{\bf B}_{{\rm p}})
= - \frac{4\pi}{c^{2}\varpi^{2}}S\frac{{\rm d}S}{{\rm d}P} \nabla P,
\en
on use of (\ref{6}) and (\ref{1}).
By (\ref{1}), (\ref{5}), (\ref{5.5}), (\ref{7}) and (\ref{8}), the
poloidal component of the force-free
equation is
\begin{equation}
\label{9}
\frac{1}{\varpi}(\nabla \times {\bf B})_{\phi}\left[1 -
\left(\frac{\alpha \varpi}{c}\right)^{2}\right]
   + \left(\frac{\alpha \varpi}{c}\right)^{2}\frac{2 B_{z}}{\varpi^{2}}
    +  \left(\frac{4\pi}{c\varpi}\right)^{2}S\frac{{\rm d}S}{{\rm d}P} =
0.
\end{equation}
Note that the first term in (\ref{9})  combines  part of the electric
force with the force due to
${\bf B}_{{\rm p}}$; the second again comes from the electric force,
while the third is due to
${\bf B}_{{\rm t}}$.

    With the sign convention in (\ref{1}), $P$ decreases from zero on the
axis, so that initially $dS/dP$ is negative, and the force density
(\ref{8}) due to the toroidal field ${\bf B}_{{\rm t}}$ acts
toward the axis. Since $S$ must vanish on the current closing
streamline, there will in general be an intermediate field line on
which $dS/dP$ and so also the volume current density ${\bf
j}_{{\rm p}}$ changes sign. Below this field line
  the force (\ref{8})
acts towards the equator. Note that the return of $S$ to zero need
not be continuous: closure via a sheet current is not excluded
(cf. Section 4). In fact at this stage, one cannot rule out that
$S$ increases monotonically from zero to its maximum on a limiting
field-streamline and then drops to zero, so that all the return
current is in the sheet. The actual allowed form for the function
$S(P)$ will emerge as part of the global solution.

    As in the earlier work, we define dimensionless coordinates
$(x,\bar{z})= (\alpha/c)(\varpi, z)$,
and normalize $P, S, {\bf E,B}$ in terms of a standard light-cylinder
field strength
$B_{{\rm lc}} = (B_{s}/2)(\alpha R/c)^{3}$:
\eq
\la{9.1}
P=\bar{P}B_{{\rm lc}}(c/\alpha)^{2},\quad S=\bar{S}B_{{\rm
lc}}(c^{2}/4\pi \alpha), \quad ({\bf B,E}) = B_{{\rm lc}}({\bf
\bar{E},\bar{B}}).
\en
(Once defined, the dimensionless quantities are again immediately
written
without the bars). The normalized
fields have the form
\eq
\la{9.2}
{\bf B}_{{\rm p}} = - \nabla P\times {\bf t}/x, \quad B_{\phi} = - S/x,
\quad  {\bf E} = \nabla P,
\en
with $P$ satisfying
\begin{equation}
\label{10}
(x^{2} - 1)\frac{\p^{2} P}{\p x^{2}} + \frac{(1+ x^{2})}{x}\frac{\p
P}{\p
x} + (x^{2} - 1)\frac{\p^{2}P}{\p z^{2}} = S\frac{{\rm d}S}{{\rm d}P}.
\end{equation}
As $(\varpi,z) \rightarrow 0$, $P$ must reduce to the normalized point
dipolar form
$-x^{2}/(x^{2} + z^{2})^{3/2}$.
The light-cylinder $x=1$ is a singularity of the differential equation
for $P$.

    An illustrative solution of (\ref{10}) within the l-c for the
non-active case $S=0$ has been discussed by Michel (1973, 1991),
Mestel \& Wang (1979) and Mestel \& Pryce (1992), and will be
referred to as the MMWP-field. The field lines which reach the
equator cross normally, forming a closed domain: the inner domain
equatorial boundary condition is \eq \la{10.5} B_{x}(x,0)=0, \en
with no equatorial current sheet. The flux function $P$ is most
easily constructed as a Fourier cosine integral (cf the Appendix).
If neither the field nor the volume current density are to be
non-singular at the l-c, (\ref{10}) with $S=0$ requires that
$B_{z} = -\p P/\p x = 0$ when $x=1$: the field lines cross the l-c
normally, and there is a neutral point at the intersection $(1,0)$
of the l-c and the equator.

   The MMWP model is of pedagogic interest through its showing how in
this
essentially relativistic problem, the macroscopic electric force
acting on the corotating GJ charge density causes a marked
deviation of {\bf B} from the curl-free dipolar form as the l-c is
approached. However, the solution is clearly incomplete, as there
is no treatment of the domain beyond the l-c. A viable, strictly
inactive model will in fact have large vacuum gaps within the l-c
(e.g. Smith et al. 2001 and references within). As discussed in
Sections 5 and 6, some features of the MMWP field should persist
in a realistic active model, such as the large dead zone, and a
similar departure from the curl-free structure near and beyond the
l-c.

\section{The domain beyond the l-c}

    Equation (\ref{4.6}) and (\ref{4.8}) combine into
\eq \la{11}
({\bf U} - \alpha\varpi {\bf t})\times {\bf B} = 0,
\en
yielding
\eq \la{11.1}
   {\bf U} = \kappa {\bf B} + \alpha \varpi {\bf t} \en
-- the sum of corotation with the star plus flow parallel to the
total field ${\bf B} = {\bf B}_{{\rm p}} + {\bf B}_{{\rm t}}$, a
result familiar from standard stellar wind theory. If ${\bf
U}_{{\rm p}}$ were to vanish beyond the l-c, then from
(\ref{4.2}), $n^+(v^+_t - \alpha \varpi) + n^-(v^-_t - \alpha
\varpi) = 0$, requiring that either the electrons or the positrons
would have to be superluminal. Thus along field lines that cross
the l-c, there is not only a net current but also a net particle
flow, as is indeed implied by the term `wind zone.' Then if, as
assumed, the simple `perfect conductivity' condition (\ref{11})
continues to hold everywhere beyond the l-c, so that there is
strictly no trans-field motion of the plasma, then in a steady
state the field must be topologically `open,' with no field lines
crossing the equator.

  A CKF-type model has the field lines crossing the equator normally
within the l-c,
as in the MMWP and FM fields, but beyond the l-c, the boundary
condition $B_{z}=0$ is imposed at the equator. In the northern
hemisphere  $B_{x} > 0$, so by (\ref{4}), near the equator the
electric field $xB_{x}$ is in the positive $z$-direction. As the
basic field is dipolar, $B_{x}$ must change sign at the equator,
implying a locally positive current density $j_{\phi}$ and a
magnetic force density that acts towards the equator. Thus the
poloidal magnetic field pinches, as in familiar non-relativistic
problems, but the combined electric force  (\ref{5.5}) and the
${\bf B}_{{\rm p}}$ force (\ref{7}) is $[(x^{2} - 1)(\nabla \times
{\bf B})_{\phi}/4\pi] {\bf k}$ and so acts away from the equator,
since $x>1$. The proposed force-free equilibrium near the equator
must therefore be maintained by the pinching effect of ${\bf
B}_{{\rm t}}$. We have seen above that $ S{\rm d}S/{\rm d}P$ does
indeed become positive at low latitudes, so that (\ref{8}) has the
required direction.

   Consider first the idealized case, with $B_{x}$, $B_{\phi}$ and
$E_{z}$  abruptly reversing sign at the equator,
and so with positive sheet currents $J_{\phi}$, $J_{x}$ and positive
surface charges $\sigma$.
Over the surface $z=+ \varepsilon$, there is a net electromagnetic
stress
$T_{ij}k_{j}$ where
\eq
\label{12}
T_{ij} = \frac{1}{4\pi}\left[-\frac{1}{2}\left({\bf E}^{2} + {\bf
B}^{2}\right)\delta_{ij}
+ (E_{i}E_{j} + B_{i}B_{j})\right]
\en
is the Maxwell stress tensor. The electric terms yield
$(x^{2}B_{x}^{2}/8\pi){\bf k}$ and the magnetic terms
$-[(B_{x}^{2} + B_{\phi}^{2})/8\pi]{\bf k}$, so the net electromagnetic
pressure, acting towards the equator, is
\eq
\label{13}
[B_{\phi}^{2} - (x^{2} - 1)B_{x}^{2}]/8\pi.
\en
There is an equal electromagnetic pressure, also acting towards the
equator on the surface $z = -\varepsilon$.
Provided
\eq
\label{14}
          |B_{\phi}/B_{x}| > \sqrt{x^{2} - 1},
\en
then an equal particle pressure at the equator is both necessary and
sufficient to maintain equilibrium.

    The force-free equation (\ref{9}) can be written succinctly as
$\partial T_{ij}/\partial x_{j} = 0$. Near
the equator, $|B_{z}| << |B_{x}|$ in this model, and the force-free
condition becomes
\eq
\label{14.5}
\frac{\partial}{\partial x_{3}}T_{33} =
\frac{1}{8\pi}\frac{\partial}{\partial z}
\left[- B_{\phi}^{2} + (x^{2} - 1)B_{x}^{2}\right] = 0,
\en
whence
\eq
\label{14.7}
\frac{1}{8\pi}[B_{\phi}^{2} - (x^{2} - 1)B_{x}^{2}] =  p_{eq}
\en
where $p_{eq}$ is independent of $z$.
If (\ref{14}) holds, then $p_{eq}$ is identified as a particle pressure
at the equator, where $B_{x}$ and
$B_{\phi}$ change sign.

   More realistically, the transition to zero field at the equator will
be
continuous, with the particle
pressure $p$ increasing steadily across half of the thin equatorial
sheet. The force-free condition
(\ref{14.5}) is replaced by
\eq
\label{14.8}
\frac{\partial}{\partial z}\left[\frac{1}{8\pi}[- B_{\phi}^{2} + (x^{2}
-
1)B_{x}^{2}] - p\right] = 0,
\en
and (\ref{14.7}) by
\eq
\label{14.9}
\frac{1}{8\pi}[B_{\phi}^{2} - (x^{2} - 1)B_{x}^{2}] + p = p_{eq}.
\en
As the equator is approached $p$ steadily increases from the uniform
value (implicit in the force-free
assumption), which could in principle be zero, to the value $p_{eq}$.
{\em Thus the model necessarily
includes a thin domain with a non-force-free electromagnetic field.}

   The limiting case, with the inequality sign in (\ref{14}) replaced by
equality,
appears consistent with zero equatorial pressure $p_{eq}$. However,
condition (\ref{9}) assumes that
all non-electromagnetic forces, including inertial forces, are small
compared with the dominant
terms in the Lorentz force.
> From (\ref{11.1}), the rotation velocity is given by
\eq \label{15}
U_{\phi} = c\left(x + \frac{U_{{\rm p}}}{c}\frac{B_{\phi}}{B_{{\rm
p}}}\right). \en

In a steady state, (\ref{4.4}) yields the energy integral in a
pressure-free system  (MRWW, MS, Contopoulos 1995)
  \eq \label{16}
\Gamma[1 - x(U_{\phi}/c)] = {\rm constant}, \en
showing that
$\Gamma$ would become infinite if $U_{\phi} = c/x, U_{{\rm p}} =
c\sqrt{(x^{2} - 1)}/x$; and from (\ref{15}) this is equivalent to
\eq \label{17}
|B_{\phi}/B_{{\rm p}}| = \sqrt{x^{2} - 1}. \en
Thus the seemingly exceptional case, with zero equatorial pressure
and so with the force-free equation holding all the way to the
equator, in fact requires that the flow outside the equatorial
zone be so highly relativistic that the neglected inertial terms
are not small, so violating an essential condition for the
force-free approximation to hold. We conclude that in all cases,
the equilibrium conditions for this model will require at least a
local breakdown in force-free conditions. In analogous
non-relativistic problems, Lynden-Bell (1996) has pointed out that
a thermal pressure is again required to balance magnetic pinching
forces, exerted locally by an otherwise force-free field. However
we again emphasize the important difference, that whereas in a
non-relativistic problem, the electric stresses are normally
smaller than the magnetic by the factor $(U/c)^{2}$, in the
present problem, beyond the l-c the opposing electric stresses
exceed the pinching poloidal field stresses by the factor $x^{2}$,
and equilibrium is possible only through the pressure exerted by
the toroidal field.

    In the domain near the equator, with $|B_{x}/B_{z}| \gg 1$, the
relativistic flow likewise has $|U_{x}/U_{z}| \gg 1$, so that
$U_{x}^{2} + U_{\phi}^{2} \approx c^{2}$, whence from (\ref{15}),
(remembering that $B_{\phi}$ is negative), \eq \label{17.5}
U_{\phi}/c = \frac{x  -  b\sqrt{(1 + b^{2} - x^{2})}}{(1 +
b^{2})}, \quad U_{x}/c = \frac{bx + \sqrt{(1 + b^{2} - x^{2})}}{(1
+ b^{2})}, \en where $b \equiv |B_{\phi}/B_{x}|$. (The
algebraically allowed choice of the opposite signs before the two
radicals would yield $U_{\phi}/c = 1$ at $x=1$, implying infinite
$\Gamma$, and so is rejected). The outflow of angular momentum
from the star in both hemispheres across a closed surface $\Sigma$
with local outward unit normal ${\bf n}$ is (e.g. Mestel 1999)
\eq \la{17.6}
  -\int(\varpi B_{\phi}/4\pi){\bf B}_{\rm p}\cdot{\bf
n}{\rm d}\Sigma =-(c/\alpha)^3B_{lc}^2\int_0^{P_c}S(P){\rm d}P,
\en
on use of (\ref{1}), (\ref{2}) and (\ref{9.1}).

\section{The $S(P)$ relation}

   As noted in Section 2, if there do exist magnetospheric models that
are
everywhere non-dissipative, with the field force-free outside
singular regions such as the equatorial sheet, then the relevant
relation $S(P)$ should emerge as part of the solution. An early
model of the whole magnetosphere by Michel (1973a, 1991) has no
dead zone, but a poloidal field that is radial all the way from
the star to infinity. In our notation, the Michel field is
\eq
\la{18} P=P_{c}\left(1\pm \frac{z}{(x^{2} +
z^{2})^{1/2}}\right),\quad B_{x} = \pm P_{c}\frac{x}{(x^{2} +
z^{2})^{3/2}}, \quad B_{z} = \pm P_{c}\frac{z}{(x^{2} +
z^{2})^{3/2}}, \en
with the negative sign applying to the northern hemisphere,
and with the critical field line $P=P_{c}$ coinciding with the
equator. From now on, just the northern hemisphere is considered.
Michel's $S(P)$ relation is
\eq \la{19}
S = -2P + \frac{P^{2}}{P_{c}} = -P_{c}\frac{x^{2}}{(x^{2} +
z^{2})},\quad B_{\phi} = - \frac{S}{x}, \en
so that
\eq \la{20}
\frac{dS}{dP} = - 2\left(1 - \frac{P}{P_{c}}\right) = -
\frac{2z}{(x^{2} + z^{2})^{1/2}}, \quad S\frac{dS}{dP} =
2P_{c}\frac{x^{2}z}{(x^{2} + z^{2})^{3/2}}. \en
One can easily
verify that (\ref{10}) is satisfied. Note that the poloidal field
(\ref{18}) is radial and (away from the equator) independent of
the spherical polar angle $\theta$, and so has no curl:
equilibrium is maintained by balance between the electric force
given by (\ref{5}) and (\ref{5.5}) and the force due to the
toroidal field (\ref{19}).  However, with the sign change at
the equator, there are again both toroidal and poloidal equatorial
sheet currents that respectively maintain the jumps in $B_{x}$ and
$B_{\phi}$; also $B_{\phi}^{2} -(x^{2}-1)B_{x}^{2} =
P_{c}^{2}/x^{4}$ when $ z=0$, so that by the above discussion,
$p_{eq} >0$.
Note also that $dS/dP = 0$ on the critical line $P_{c}$, but is negative
for
$0< P/P_c<1$: in the Michel model, all the return current is in the
equatorial
sheet.

   In the generalized CKF picture, with the poloidal field forced to have
a dipolar rather than a radial structure near the star, there is an
associated dead zone within the l-c, similar to that found in the MMWP,
FM and MS fields, and analogous to that in the non-relativistic wind
problem (e.g. Mestel and Spruit 1987).
The dead zone terminates at the point $(x_c , 0)$; the value of
$x_c (\leq 1)$ will be seen to be an extra parameter, fixing
the global field structure. Within the dead zone the field lines
close, crossing the equator normally
so that the appropriate equatorial boundary condition is $P_z \propto
B_x(x,0) = 0$ for $x< x_c$.
The dead zone is bounded by the separatrix field line $P_c$. In the wind
zone outside of $P_c$,
the wind flow is along the poloidal field, and so the equatorial
boundary
condition is $P_x \propto B_z
= 0$, but now extending through the l-c from $x_c$ to $\infty$, with
again a
charge-current sheet maintaining the discontinuous sign changes in
$E_{z}, B_{x}$ and
$B_{\phi}$. Thus the critical field line $P(x,0) =  P_{c}$
extending along the equator is the continuation of the separatrix
between
the wind and dead zones.

   Just outside the equatorial charge-current sheet and the separatrix,
(\ref{10}) holds all the way in from $\infty$. At the intersection
(1,0) with the l-c and just outside the sheet, with $P_{x} \propto
B_{z}= 0$ and no singularities in $P_{xx}$ and $P_{zz}$,
(\ref{10}) requires that the constant value of $SdS/dP$ along
$P_{c}$ must be zero. But $S(P_{c}) \neq 0$, because as seen
above, beyond the l-c the pinching force exerted by $B_{\phi}$ is
necessary for equilibrium; we therefore require, as in the Michel
field,
\eq\la{21}
            \left(\frac{dS}{dP}\right) = 0\en
on the critical field line $P_{c}$. This condition is propagated inwards
along $P_c$ on to its continuation,
the separatrix $P_{c}$ between the wind and dead zones:  the equatorial
equilibrium conditions beyond the l-c
impose a constraint on the global $S(P)$ relation. From (\ref{6}),
condition (\ref{21}) requires
that the poloidal volume current density ${\bf j}_{{\rm p}}$ falls to
zero on $P_{c}$,
but the finite value for $S(P_{c})$ requires a poloidal current sheet at
the equator beyond $x_c$ and along the separatrix within it.


\section{The domain within the light-cylinder}

     We have emphasized that beyond the neutral point $(x_c,0)$,  there
is a finite thermal pressure on the equator, necessary for
equilibrium. Likewise, it is not obvious that within the l-c, the
boundary condition on the field line separatrix between the wind
zone (labelled 2) and the dead zone (labelled 1) can be satisfied
without a thermal pressure within the dead zone (see Fig. 1).

\begin{figure}
$$\vbox{\psfig{figure=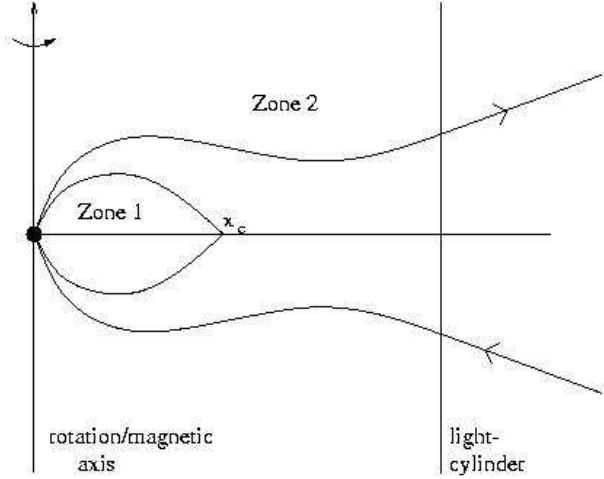,width=8.7truecm,angle=0}}$$
\caption{Schematic magnetosphere}
\label{Fig 1}
\end{figure}

  It is in fact easy to generalize (\ref{9}) and (\ref{10}) to include a
pressure gradient: with gravity and inertia still negligible,
\eq
\la{22}
-\frac{\nabla P}{4\pi}\left\{ \frac{1}{\varpi}(\nabla \times {\bf
B})_{\phi}
\left[1-\left(\frac{\alpha \varpi}{c}\right)^{2}\right]
   + \left(\frac{\alpha \varpi}{c}\right)^{2}\frac{2 B_{z}}{\varpi^{2}}
    +  \left(\frac{4\pi}{c\varpi}\right)^{2}S\frac{{\rm d}S}{{\rm
d}P}\right\} - \nabla p = 0.
\end{equation}
Thus $p=p(P)$ --  the constant pressure
surfaces must coincide with the poloidal field lines. In
normalized form, (\ref{22}) becomes
\eq
\la{23}
(1-x^{2})\frac{\p^{2}P}{\p x^{2}} - \frac{(1+x^{2})}{x}\frac{\p P}{\p x}
+ (1-x^{2})\frac{\p^{2}P}{\p z^{2}} = -S\frac{{\rm d}S}{{\rm d}P}
-\frac{1}{2}x^{2}\frac{{\rm d}p}{{\rm d}P}.
\en

{\em Prima facie}, there is no obvious objection to the adoption of the
simplest case,
with $p$ the same  constant on all the field lines within the dead zone
and zero in the wind zone, i.e. with a discontinuity in both $p$ and
$S(P)$ on
the separatrix. The equation for $P$ within the wind
zone then remains (\ref{10}) (with signs reversed for convenience):
\eq
\la{24}
(1 - x^{2})\frac{\p^{2}P}{\p x^{2}} - \frac{(1 + x^{2})}{x}\frac{\p
P}{\p
x} + (1 - x^{2})\frac{\p^{2}P}
{\p z^{2}} = - S\frac{{\rm d}S}{{\rm d}P};
\en
and in the dead zone we have the same equation with $S=0$.
However, along the separatrix $P_{c}$, extending inwards from the
equatorial point
$(x_c,\, 0)$, the equilibrium conditions require a discontinuity
in $B_{{\rm p}}$ as well as those in $p$ and $S(P)$.
Writing
\eq
\la{25}
n_{i} = -[({\bf t} \times ({\bf B}_{{\rm p}}/B_{{\rm p}})]_{i}
\en
the unit normal to the separatrix, we require continuity of
\eq
\la{26}
[-(8\pi p + E^{2} + B_{{\rm p}}^{2} + B_{\phi}^{2})\delta_{ij} +
2E_{i}E_{j} +
2B_{i}B_{j}]n_{j}
\en
with ${\bf E}$ given by (\ref{4}).
The components of (\ref{26}) parallel to ${\bf t}$ and to the separatrix
are automatically zero.
The component normal to $P_{c}$ reduces to continuity of
\eq
\la{27}
8\pi p + B_{{\rm p}}^{2}[1 - (\alpha \varpi/c)^{2}] + B_{\phi}^{2},
\en
i.e. to
\eq
\la{28}
8\pi p + B_{{\rm p}1}^{2}(1 - x^{2}) = B_{{\rm p}2}^{2}(1 - x^{2}) +
B_{\phi2}^{2}.
\en
It is convenient to normalize $p$ in units of $B_{{\rm lc}}^2/8\pi$.

   For $x>x_c$, the wind zone 2 extends from the equator $z=0$ to $z=
\infty$;
for $x < x_c$, zone 2 extends from $z_c(x)$ defined by the separatrix
\eq
\la{28.1}
  P(x, z_c) = P_c
  \en
  to $z = \infty.$
The separatrix function (\ref{28.1}) is not known {\em a priori}
but must emerge as part of the solution by iteration. We assume
provisionally that at the point $(x_c, 0)$ (referred to as N), the
poloidal field $B_{{\rm p}} =0$ both just outside and just inside
the separatrix. From (\ref{28}), the pressure of the toroidal
field is balanced by the thermal pressure $p$, so that
\eq
\la{28.5}
  p = S^2(P_c)/x_c^2. \en
  For $x<x_c$, with use of
(\ref{9.1}) and (\ref{9.2}), (\ref{28}) then becomes
\eq \la{28.6}
(\nabla P_{1})^{2} = (\nabla P_{2})^{2} + S^{2}(P_{c})\frac{(1 -
x^{2}/x_{c}^{2})}{(1-x^{2})}. \en
The discontinuity in $|\nabla
P|$ grows from zero as $x$ moves in from $x_c$ but will become a
small fraction of $|\nabla P|$ for small $x$.

    When $x_c<1$, the function $P$ has a simple
analytical behaviour near the neutral point $(x_c,\,0)$. The separatrix
leaves
$(x_c,\,0)$
  making an angle $\theta = 2\pi/3$ with the outward-pointing equator. To
leading order, in both the dead and wind zones, $P_{xx} + P_{zz} = 0$,
which has
the local solution
\eq
\la{28.7}
P/P_c  = 1 + A_{2,1}R^{3/2}\sin(3\theta/2),\,\,\, R^2 = z^2 + (x-x_c)^2,
\en
where the coefficients $A_{2,1}$ apply respectively to the wind zone
$0 < \theta < 2\pi/3$ and the dead zone $2\pi/3 < \theta < \pi$. It is
seen that
along $\theta = 0$, $P= P_c$, and on $\theta = \pi$, $P_{\theta} \propto
P_z = 0$,
as required. Across the separatrix $z=\sqrt{3}(x_c - x)$, $P$ is
continuous, while the jump condition (\ref{28.6}) then yields
\eq \la{28.8}
A_1^2 = A_2^2 + \frac{4S^2(P_c)}{9P_c^2 x_c(1 -
x_c^2)}. \en

  At this point, it is instructive to make a comparison with the
analogous
non-relativistic problem, in which the electric stresses are small
by factors $O(v/c)^{2}$ and so are negligible, and also inertial
forces are still neglected. Suppose that there is again a
separatrix passing through the poloidal field neutral point at
$(x_c, \,0)$. The balance equation across the separatrix is now
continuity of $8\pi p + B_{{\rm p}}^{2} + B_{\phi}^{2}$ -- the
$(1-x^2)$ factors in
(\ref{28}) are replaced by unity. If again $p$ is negligible in
the wind zone, then the constant value of $p$ along  the
separatrix within the dead zone is again fixed by the condition at
the neutral point: $p = S^{2}(P_{c})/x_c^2$ in normalized form.
The balance condition is then $S^{2}(P_{c})(1 - x^{2}/x_c^2) +
(\nabla P_{2})^{2} = (\nabla P_{1})^{2}$. The presence of the
factor $(1-x^{2}/x_c^{2})$ in the $S^{2}$ term enables ${\bf
B}_{{\rm p}}$ to be continuous (and zero) at the neutral point
$(x_c,\,0)$, but with $S^2(P_c)$ still non-zero. (Clearly, in the
non-relativistic problem, the numerical value of $x_c$ is of no
significance.)

   By contrast, in the relativistic problem, near the l-c the electric
field strength
approaches the poloidal magnetic field strength, so that the factor
$(1-x^{2})$ now appears
multiplying both the $(\nabla P)^{2}$ terms in (\ref{28.6}).
If now one were to take  $x_c = 1$, then
the non-vanishing factor $(1-x^{2})$ would cancel, and (\ref{28.6})
would
reduce to
\eq
\la{29}
  S^{2}(P_{c}) + (\nabla P_{2})^{2} = (\nabla P_{1})^{2}.
\en
along the separatrix. At the point $(x_c,\,0)$, the simultaneous
vanishing of $\nabla P_1$ and $\nabla P_2$ would then require that
$S(P_c)=0$ (and
so also by (\ref{28.5}) $p =0$). But if beyond the l-c the equatorial
toroidal component $B_{\phi} = - S(P_c)/x$ were zero, then the balance
condition could not be
satisfied (cf. (\ref{14.7})).

   In fact, as discussed in the Appendix, the possible vanishing of
$S$ on the separatrix is appropriate to models with a radically
different external equatorial boundary condition. In the present
problem, the case $x_c=1$, which yields (\ref{29}), may
be incorporated provided that for this limiting case, we allow
$\nabla P_1 \neq 0$, i.e. the discontinuity in $B_{{\rm p}}$
across the separatrix persists at the equator. If $x_c = 1 -
\epsilon$ with $\epsilon \ll 1$, then from (\ref{28.6}), the
equatorial $B_z$ can be zero on both sides of the separatrix,
without requiring that $S(P_c) = 0$. However, at a neighbouring
separatrix point $x= x_c - X$ with $X \ll 1$,
  the second term in (\ref{28.6}) will
have climbed from zero at N to $S^2(P_c)[X/(X+\epsilon)] \simeq
S^2(P_c)$ once $X\gg\epsilon$. If solutions with $S(P_c) \neq 0$
continue to exist as $x_c \to 1$, then the separatrix balance
condition will imply a steeper and steeper local gradient in
$B_z$, indeed tending to a discontinuity when $x_c = 1$.

  For the limiting case $x_c=1$, an appropriate local model, valid near
the critical point $x=1,\,z=0$, has $|\nabla P|^2$ jumping by a
constant value across the separatrix
  $\bar{P} \equiv P/P_c = 1$. New independent variables $(r,t)$ and the
dependent
variable
$u$ are introduced, defined by
\eq \la{29.1}
  x = 1  - r\sin t,\quad z=r\cos t,\,\quad r^2 = (x-1)^2 + z^2;\quad\quad
\bar{P} = 1 + u. \en When $r$ and $u$ are small, (\ref{24})
reduces to \eq\la{29.2} (1 -x^2)(u_{xx} + u_{zz}) - (1 + x^2)u_x/x
= 0 \en
  in the dead zone. In the contiguous wind zone , the right-hand
  side is again $SS^{\prime}$, but as this will behave like $u$,
the form (\ref{29.2}) is appropriate on both sides of the
separatrix. To leading order, this reduces to
\eq\la{29.3}
  u_{rr}+ 2\frac{u_r}{r} + \frac{1}{r^2 \sin t}[\sin t u_t]_t =0.
\en
The appropriate boundary conditions are:\\
(1) $u_t = 0$ on $t = \pi/2$ -- normal crossing of the equator;\\
(2) $u=0$ on the separatrix, leaving the critical point at the as
yet unknown angle $t=t_0$;\\
(3) $(u_t/r)^2$ jumps by a constant on $t=t_0$.\\
The solution of (\ref{29.3}) satisfying these conditions is:
\begin{eqnarray}
\la{29.4}
&{\rm Domain\,\, a}& \quad 0 < t < t_0: \quad \quad \quad \quad
u=0,\nonumber \\
&{\rm Domain\,\, b}& \quad t_0 < t < \pi/2: \quad \quad \quad u =
ArQ_1[\cos t],
\end{eqnarray}
where $Q_1(X)$ is the Legendre function $Q_1 = - 1 +
(X/2)\log[(1+X)/(1-X)]$, and $t_0$, given by $Q_1(\cos t_0) = 0$,
is $33.5342^\circ$. The constant $A$ then follows from $u_t/r = S(1)$.
In Domain b, $Q_1[\cos t] = -1 - \cos t\log(\tan t/2)$, and $\p
Q_1/\p(\cos t)= \cos t/\sin^2t - \log(\tan t/2)$; whence at points
near the l-c ($x$ close to unity),
\eq \la{29.5}
B_x \propto \p u/\p z = -A\log(\tan t/2), \quad\quad
B_z \propto -\p u/\p x = A/\sin t, \en
yielding $B_x=0, B_z \propto A = {\rm constant}$ on the equator.

   An essential step in all the argument is the condition $p = p(P)$,
following from (\ref{22}).
If instead the thermal pressure $p$ were (illicitly) allowed to vary so
as to balance the pressure $B_{\phi2}^{2} = S^{2}(P_{c})/x^{2}$
exerted by the external toroidal field, then from (\ref{28}) there would
be no discontinuity in $B_{{\rm p}}$ at the equator or indeed anywhere
along the separatrix.

It is appropriate also to re-emphasize just how crucial to the
discussion is the outer domain equatorial boundary condition
\eq\la{29.05}
B_{z}(x,0) = 0 \en
  for $x > 1$;
for it is the consequent local equilibrium requirement $S(P_{c}) > 0$
that implies a
poloidal sheet current within the l-c along the separatrix, leading to
the conditions
({\ref{28}) and (\ref{28.6}), with a non-vanishing $B_{\phi2}$. It is
then clear that
a non-zero $p(P_c)$ is required for the balance condition to hold at the
neutral point.

\section{Model construction}

  The two basic parameters of a pulsar are clearly the angular velocity
and the
dipole moment of the neutron star. The spin-down rate of an active
pulsar
will
depend on the strength of the circulating poloidal current. In the
Michel
$S(P)$
relation (\ref{19}), the form with $S\simeq -2P$ valid near the poles
derives from taking the current density flowing along the axis as
equivalent to the G-J
electron charge density moving with the speed $c$, a reasonable upper
limit, and indeed
one that must be closely approached if the electrons are to generate an
electron-positron
pair plasma near the star (cf. e.g. MS, Section 3). We
consider only $S(P)$
relations that behave like the Michel form for small $P$.

   The only other parameter
introduced into the theory is the pressure in the dead zone, which we
have taken as uniform.
{\em Prima facie}, there should be
a class of possible functions $S(P)$, each fixing the global
field function $P$ and simultaneously the separatrix parameter $P_c$ and
the
limit $x_c$ of the dead zone, with the value of the required dead zone
pressure $p(P_c)$ given by (\ref{28.5}).
Equivalently, a choice of $x_c$ should have an associated class of
functions
$S(P)$, each fixing the global $P$, the value of $P_c$ and of $p(P_c)$,
and the shape of the separatrix $P(x,z)=P_c$ between the star and $x_c$.
  Intuitively, one expects an increase
in $p(P_c)$  to correspond to a decrease in  $x_c.$

    The functions $P(x,z),\, S(P)$ are renormalized in terms of the
  value $P_c$ pertaining to the separatrix and its extension along the
equator
  $z=0$, extending from the point $x_c <1$, through the light-cylinder
$x=1$ to $\infty$:
\eq
\la{29.7}
\bar{P} = P/P_c, \quad \quad \bar{S} = S/P_c.
\en
  Although by convention the unnormalized $P$ is negative, the new
normalized form $\bar{P}$ is positive, and
$\bar{S}$ is negative. (From now on the bar will be dropped, all
quantities being assumed renormalized.) On the separatrix and its
equatorial continuation, the normalized $P$ is unity.
The boundary condition at the origin then becomes $P\simeq
(-1/P_c)x^2/(x^2 + z^2)^{3/2}$.

  The equation for $P$ is
  \eq
  \la{30.1}
   (x^2 - 1)P_{xx} + \frac{(1+ x^2)}{x}P_x  + (x^2 - 1)P_{zz} =
S\frac{dS}{dP}.
  \en

Defining the spherical polar coordinates $(r,\theta)$ by
$z=r\cos\theta$,
$x=r\sin\theta$, it is to be expected that $P\sim r^n f(\theta)$ as
$r\to\infty$, for some power $n$ and function $f$. Such a homogeneous
form
is consistent with the equatorial condition $P=1$ on
$\theta={{\textstyle
{1\over2}}}\pi$, only if
$n=0$, giving a radial field. The appropriate boundary condition at
infinity is thus $\partial P/\partial r=0$. For large $r$, (\ref{30.1})
takes the form
$x^2\nabla^2P=SS'(P)$ which  integrates once to require
\eq
\la{radialB}
\sin\theta {dP\over d\theta}=S(P)\qquad\mbox{imposing\quad S(0)=0.}
\en
For the Michel field $P=1-\cos\theta$ and $S=2P-P^2$, but a solution
consistent with the dipole at the origin has a different function
$S(P)$. It follows that
the radial field at infinity must differ from the Michel form. It is
however
found in the next section that the difference is not so great.

\subsection{Numerical Method}

The function $P(x,\,z)$ satisfies an elliptic, quasilinear
equation with a mixture of Neumann, Dirichlet and Robin boundary
conditions. The PDE contains an unknown function $S(P)$ which must
be determined by a regularity condition across the singular line
$x=1$. There is a discontinuity in normal derivative across the
curve $P=1$ whose position is unknown {\em a priori}. The solution
is driven by the dipole of strength $1/P_c$ at the origin. A
further numerical difficulty derives from the non-analytic
behaviour of the solution near the point $(x_c,0)$.

The problem is determined by the two parameters $P_c$ and $p$, but
for convenience solutions are sought for fixed $P_c$ and $x_c$.
This reduces the amount by which the curve $P=1$ requires
adjustment during the solution.

Equation (\ref{30.1}) is discretised using a finite difference
scheme on a rectangular grid $(m\delta x,\,n\delta z)$ for $1\leq
m\leq M$ and $0\leq n\leq N$, and $P_{mn}$ denotes the sought
approximation to $P$ at the grid point. The l-c is at $m=m_0$, so
that $m_0\delta x=1$, and the computational domain is $0<x<x_m$,
$0<z<z_m$ where $x_m=M\delta x$ and $z_m=N\delta z$.

The position of the dead-zone boundary $P=1$ is marked by its
intersections with the grid lines. Away from this boundary and for
$m\not=m_0$, second order centred difference approximations are
used for $P_x$, $P_{xx}$ and $P_{zz}$, regarding $SS'(P)$ as a
known source term. In fact, the dipole at the origin is subtracted
out from $P$ before differencing, and the dipole is differentiated
analytically.

  Equations for $P_{m_0n}$ and for $P_{mn}$ near the curve $P=1$
are derived separately as discussed below. The resulting equations
are written in a form appropriate for Jacobi iteration. Thus at
each iteration new values of $P_{mn}$ are derived from the old
values, for given $S(P)$ and position of the curve $P=1$. Once the
new values of $P_{mn}$ are found, values for $SS'(P)$ are found
using
$$SS'(P_{m_0n})=(P_{m_0+1\,n}-P_{m_0-1\,n})/\delta x.$$
These values are then interpolated using splines to give the new
function $SS'(P)$, which is integrated to give a new value of
$S(1)$. For values of $P$ smaller than those reached along the
finite length $0<z<z_m$ the Michel solution was used for $S(P)$.
The intersection points of the curve $P=1$ with the grid lines are
then updated by requiring the jump in $|\nabla P|^2$ to be
correct. The internal and external values of $|\nabla P|$ are then
used on the next iteration when approximating the derivatives at
the grid points close to $P=1$. The point $(x_c,\,0)$ is kept
fixed during the iteration, and a local balance there requires the
system to converge to suitable values of $p$ and $S(1)$.

Assuming a smooth crossing of the l-c, $P(x,\,z)$ is expanded near
$x=1$ as a power series in $(x-1)$, which is substituted into
(\ref{30.1}). This gives the results
\eq \label{lceqn}
SS'(P)=2P_x\qquad\hbox{and}\qquad 4P_{xx}+2P_{zz}=-(SS')_x \en
on $x=1$. This latter equation is discretised and used for $m=m_0$
in place of (\ref{30.1}). Finally, on the edge of the domain, the
conditions $P(0,\,z)=0$, $P_z(x,\,0)=0$ for $x<x_c$ while
$P(x,\,0)=1$ for $x>x_c$, and $xP_x+zP_z=0$ on $x=x_m$ and on
$z=z_m$ are imposed.

The above equations are solved iteratively. The numerical
behaviour is improved using continuation techniques from a nearby
solution, and by the use of under-relaxation.
As well as the fundamental parameters $P_c$ and $x_c$, the domain size
$x_m$ and $z_m$ and the grid size $\delta x$ and $\delta z$ affect the
numerical solution. These latter values can be varied to ensure the
computation is robust.

No difficulties are encountered in the vicinity of the l-c nor the
dipole. However, when the `front' $P=1$ crosses one of the grid
points in the course of the iteration some readjustment occurs and
the convergence is slower.
Similar effects occur in the numerical solution of Stefan problems with
a liquid-solid
interface.

Solutions are found for some but not all parameter pairs
$(P_c,\,x_c)$. For a fixed value of $x_c$, the scheme converges
only if $P_c$ lies in some interval $P_1(x_c)<P_c<P_2(x_c)$. As
$x_c\to1$, the values of $P_1$ and $P_2$ decrease, and become
proportionally closer.
If $P_c$ is too small the dead-zone becomes non-convex, bulging towards
the l-c away from the equator.
It is believed that solutions cease to exist once $P$-lines attempt to
cross the l-c three times rather than once.
If $P_c$ is too large, the curve
$P=1$ reaches the equator at a value of $x<x_c$. Some dead-zone
$P$-lines then
cross the equator in the interval $(x_c,\,1)$, and are not linked with
the
star and such solutions are deemed irrelevant.

\begin{figure}
$$\vbox{\psfig{figure=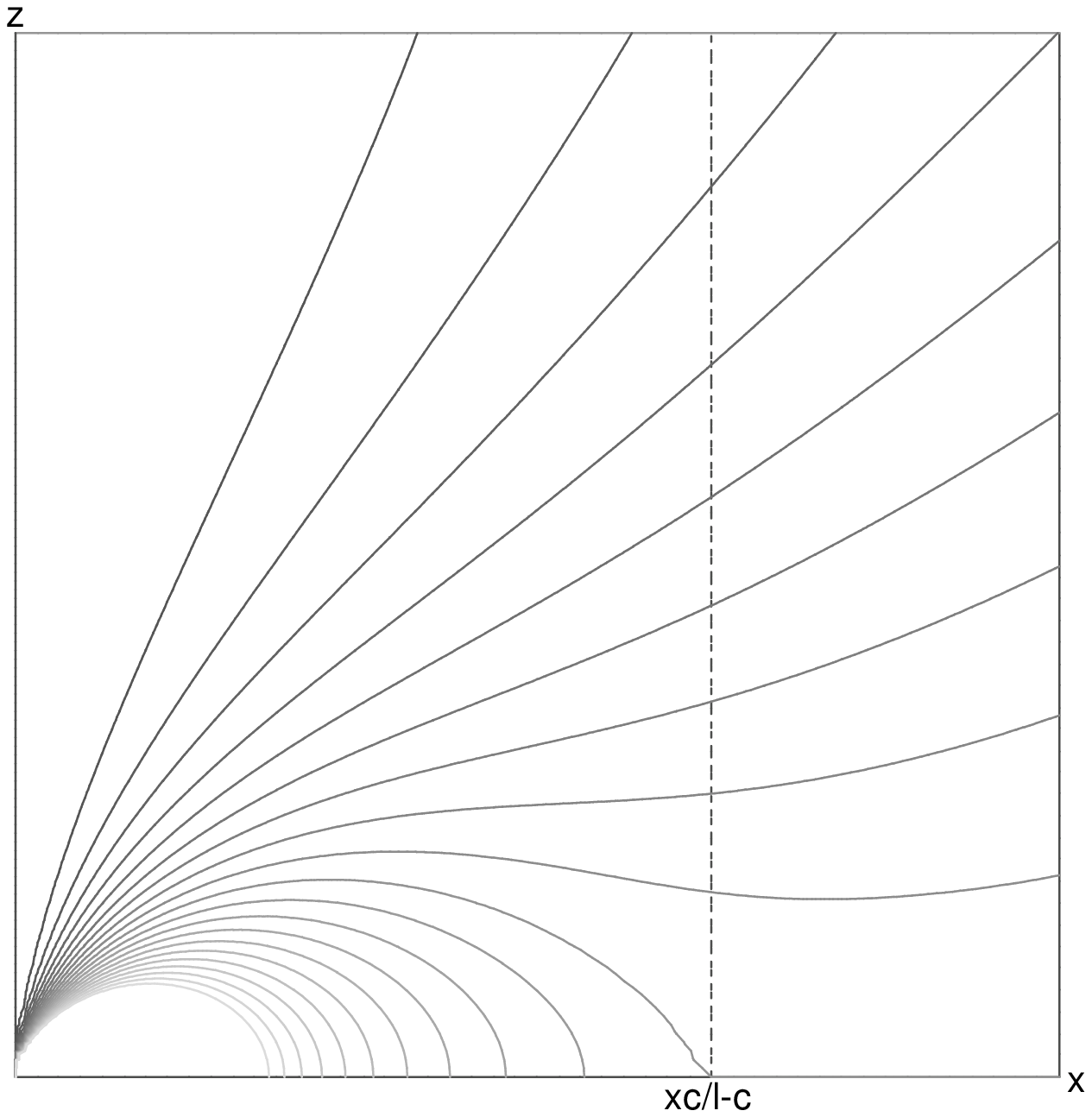,width=4.5truecm,angle=270}}\qquad
\vbox{\psfig{figure=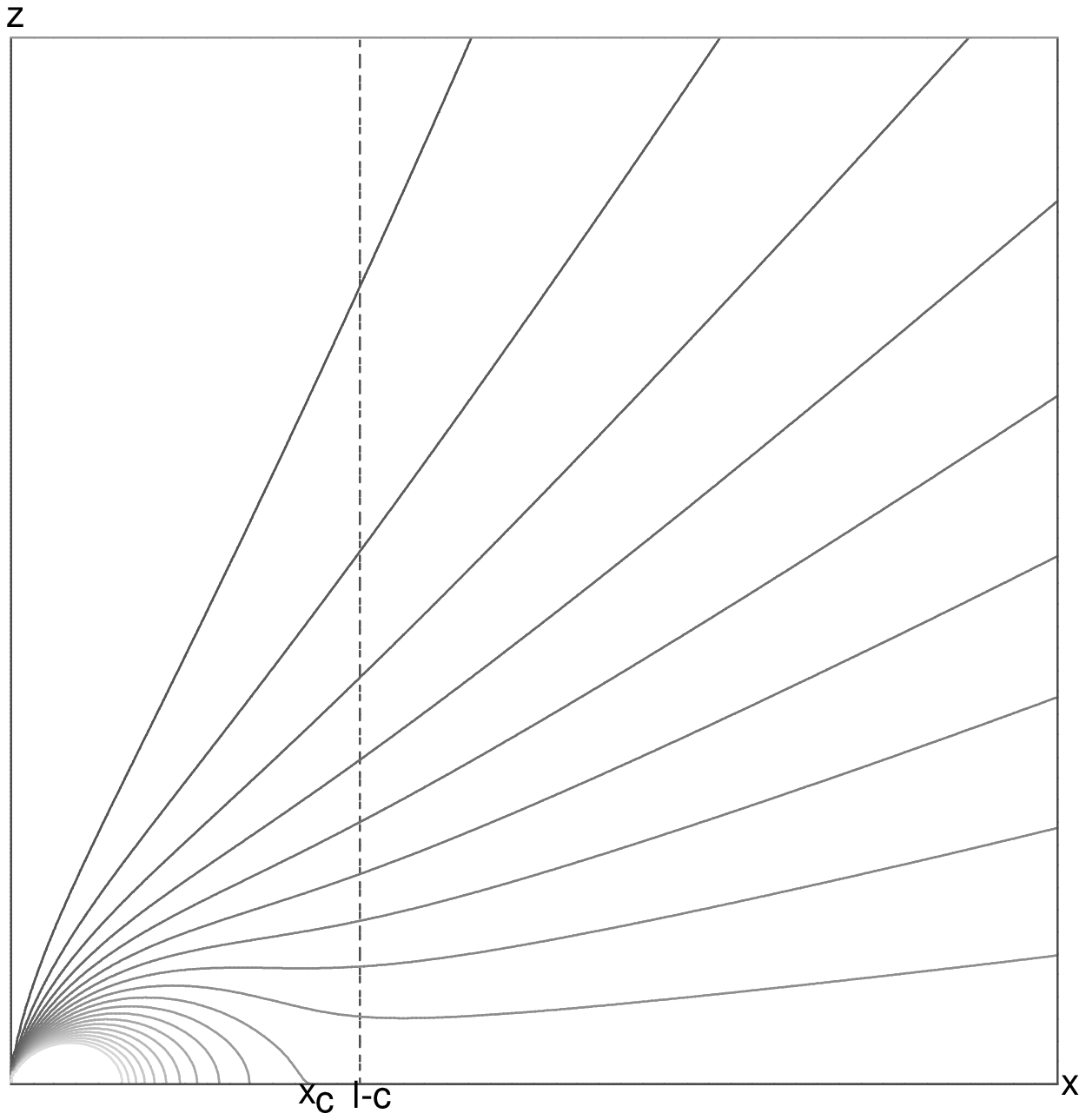,width=4.5truecm,angle=270}}\qquad
\vbox{\psfig{figure=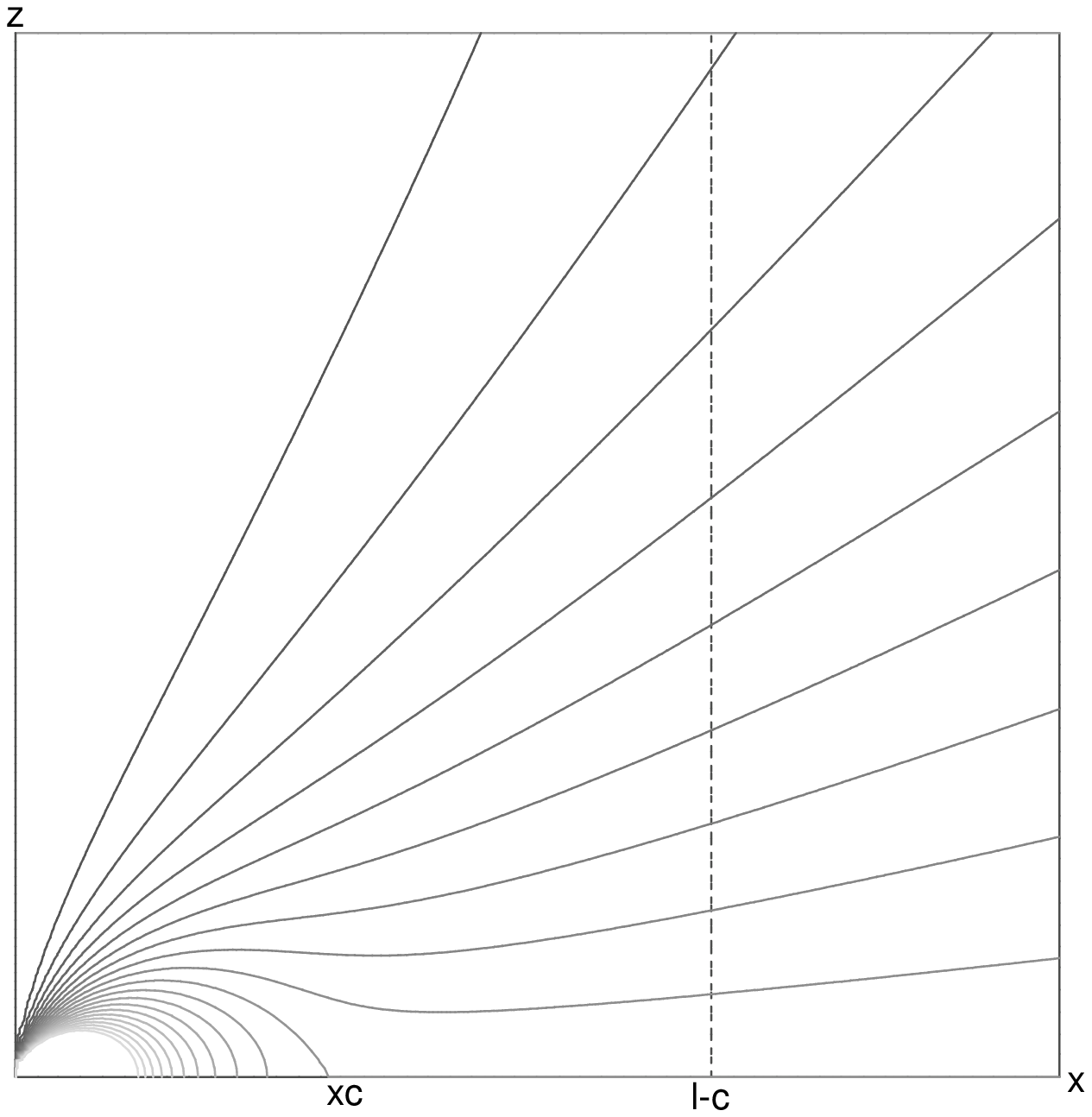,width=4.5truecm,angle=270}}$$
\caption{Contours of $P(x,z)$ Left: $x_c=1$, $P_c=1.46$; Middle:
$x_c=0.86$, $P_c=1.65$;
Right; $x_c=0.45$, $P_c=2.9$.}
\label{figP}
\end{figure}

The left diagram in figure \ref{figP} portrays $P(x,\,z)$ in the
limiting case
$x_c=1$ with $P_c=1.46$, for which it is found $S(1)\simeq0.919$.
An approximately linear variation with distance from $(1,\,0)$
within the dead zone is found, in keeping with (\ref{29.4}). With
step-lengths $\delta x=\delta z=0.025$, the value of $P(1-\delta
x,0)$ predicts a value of $A$ in (\ref{29.4}) in agreement with
the theoretical estimate to two significant figures.

Solutions with $x_c<1$ are similar in appearance, although the
local structure around the point $(x_c,\,0)$ differs as discussed
above. It was not possible to distinguish the predicted separatrix
angles $30^\circ$ of (\ref28.7}) and $33^\circ$ of (\ref{29.4}).
The centre of figure \ref{figP} shows the case $x_c=0.86$
$P_c=1.65$ for a larger domain, while the right depicts a case
with a weaker dipole $P_c=2.9$ and smaller dead zone $x_c=0.45$,
with equivalently an increase in the poloidal flux extending to
infinity.

\begin{figure}
$$\vbox{\psfig{figure=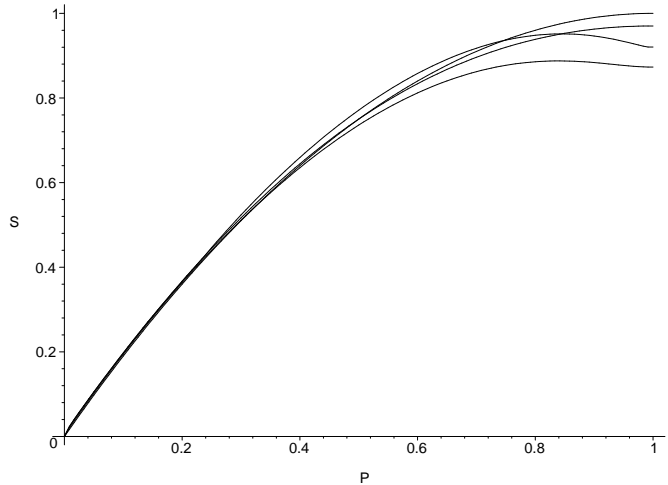,width=8.7truecm,angle=270}}$$
\caption{$S(P)$, for the cases in figure \ref{figP}. Bottom to top:
$x_c=0.86,\,1,\,0.45$, Michel.}
\label{figSP}
\end{figure}

The corresponding $S(P)$ variations are plotted in figure
\ref{figSP}, along with the quadratic Michel relation. For $x_c$
some distance from the l-c $S(P)$ is monotonic and the field on
the l-c diverges from the equator.  As $x_c$ approaches unity, a
region of negative $S'(P)$ develops near $P=1$, indicating that on
the l-c $P_x$ becomes negative so that the field is converging
towards the equator, although it becomes radial further out. The
functional form of $S(P)$ never seems to differ wildly from the
Michel solution, but $P(x,\,z)$ is affected substantially. This
supports an observational point made at the outset of this paper,
that magnetosphere currents may need very little alteration to
achieve a different field configuration, in particular changes in
the polar cap radius. Our solutions here correspond to polar caps
which are respectively 20\% ($x_c=1$), 30\% ($x_c=.86$) and 70\% 
($x_c=.45$) greater than
the vacuum dipole case conventionally assumed in interpreting pulsar
profile widths.

\section{Discussion}
\subsection{The equatorial boundary condition}
   After considerable effort we have amended the pioneering study of CKF
and constructed a class of models subject in the wind zone to the
equatorial boundary condition $B_z = 0$, i.e. with field lines in
the wind zone never crossing the equator, as is indeed required by
strict application of the `perfect conductivity', dissipation-free
condition.
It has been emphasized that although the constructed magnetospheric
field
is force-free nearly everywhere, it is non-force-free in the the
pinched equatorial zone where the field changes direction, and
also along the separatrix between the wind and dead zones. Also,
in general, as found earlier in CKF, the current returns to the
star partly as a volume current, and partly as a sheet current
along the equator and the separatrix.

  By definition, the global force-free field condition describes an
{\em electromagnetically} dominant system. In non-relativistic
MHD, one would associate a poloidal field, drawn out so as to have
a predominant $x-$component, rather with a {\em
kinetically}-dominant system, with the Reynolds stresses at least
locally stronger than the Maxwell stresses. Without such an
outward pull, an equatorially pinching, quasi-radial field is
likely to be unstable against recurrent reconnection, spontaneously
converting into a structure with field lines closing across the
equator.

  Recall that in the analogous stellar problem (e.g. Mestel
and Spruit 1987, Mestel 1999, 2003), it is argued that the
Alfv\'enic surface, on which the magnetic and kinetic energy
surfaces are comparable, not only defines precisely the extent of
`effective
corotation,' but is also a rough demarcation
between the zones with respectively closed and open field lines.
Suppose now
that in studying the relativistic force-free equation, we
consider models with a domain in which some field lines cross the l-c
and
close beyond it. Gas in this domain cannot corotate with the star,
but will flow into the equatorial zone. A hypothetical steady state will
clearly require that the accumulating gas be able to flow outwards,
crossing
the field through a macro-resistivity.

  The most extreme example will have these
field lines approaching the equator normally, so that the
equatorial boundary condition $B_x=0$, holding within the l-c, also
replaces
$B_z=0$, for some way beyond the l-c. However, there is
a further significant difference from the non-relativistic problem:
a state with the equatorial boundary condition $B_x = 0, B_z
\neq 0$ appears to be {\em electromagnetically} unstable. If the
field is perturbed so that there appear components $\pm B_x$ on
the $\pm z$ surfaces of the equatorial sheet, then the Maxwell
stresses yield $x$-components $(E_xE_z + B_xB_z)/2\pi$; and
provided coupling with the rotating magnetic star is maintained,
this becomes $(1 - x^2)B_xB_z/2\pi$. Since $B_z < 0$, when $x >
1$, the force acts to drive the frozen-in gas in the direction of
$B_x$, so increasing the imposed perturbation. The effect is again
due to the electric part of the stress, which beyond the l-c dominates
over the poloidal magnetic field contribution.
Thus there may exist steady states, alternative to those studied in
this paper, in which the field at the equator is neither parallel
nor perpendicular to the plane, but approaches it obliquely, and with
the
equatorial gas driven out by a combination of electrical and centrifugal
force, and crossing the $B_z$ field component via a dynamically-driven
macro-resistivity.

   Any such change in the equatorial boundary condition will react
on the global solution of the force-free equation (cf. the
Appendix). In MS (1994), two cases were studied, both subject to
the special equatorial condition $B_x=0$: (a) the MMWP dead model,
with $S=0$, extrapolated beyond the l-c; and (b) the special live
model, with $S = -2P + 2P^2/P_d$, where $P_d$ is the separatrix
bordering the dead and wind zones. Note that in this tentative
live model, $S(P_d)=0$: all of the current returns as volume
current, with no current sheet on the separatrix. In both these
cases, it was found impossible to construct a solution that was
smoothly continuous on the l-c, and that extended to infinity: the
solutions became singular at $x=2$ and $x=1.4$ respectively (cf
Mestel 1999 and 2003, p. 547 and p. 612).

  For the second case, it was proposed that there exists near $x=1.4$
a thin volume domain (taken as cylindrical) in which ${\bf E\cdot B}
\neq 0$. A possible physical reason for the effective resistivity
is the MHD instability of a local dominantly toroidal field 
(e.g. Begelman 1998). Prima
facie, this allows construction of a solution that is well-behaved
at both the l-c and infinity.
  However, it is possible that the failure to find a global solution
of the dissipation-free equation might turn out to be just a
consequence of the assumed absence of a current sheet on the
separatrix. This remains a problem for the future, along with a
possible generalization to the cases with oblique approach of the
field to the equator. A fully convincing treatment will require
study of the gas and current flow in the equatorial zone.

\subsection{The inertial terms}

The solutions constructed according to the procedure of Section 6
may be regarded as appropriate generalizations of the Michel
field, modified by imposition of the dipolar field on the star,
with its associated dead zone. The parameters of the problem are
assumed to yield the electromagnetic energy density dominating
over the kinetic energy density, so that the deviation from
force-free conditions is restricted to the thin equatorial domain.
As the outflowing gas will be accelerated to high $\Gamma$-values,
we need to check that the inertial terms remain small far beyond
the l-c.

    Refer back to the equations (\ref{4.3})-(\ref{4.7}). We
are interested now in the 
parameter domain in which the ideal MHD condition (\ref{4.6}) remains an 
adequate approximation to (\ref{4.5}), but the non-linear inertial term in 
(\ref{4.4}) may no 
longer be negligible. We follow the standard treatment of a perfectly 
conducting, relativistic wind, flowing in the presence of the asymptotically
radial Michel field and the associated toroidal field (Michel
1969; Goldreich and Julian 1970; Li and Melrose 1994; Mestel 1999,
Section 7.9), using the notation of Melatos and Melrose. In
addition to the kinematic relation (\ref{11.1}), the continuity
condition (\ref{4.3}) has the steady state integral
  \eq\la{43}
  mn\kappa = \eta \en
with $m$ the electron/positron mass, and the equation of motion (\ref{4.4}) 
yields the modified torque
and energy integrals
\eq \la{44}
-\varpi B_{\phi}/4\pi + \eta
\Gamma\varpi U_{\phi} = - \beta(P)/4\pi \en
and
\eq \la{45}
  \Gamma c^2(1 - \alpha \varpi U_{\phi}) = H(P).
  \en
(Recall that in this relativistic problem the electric force density 
$\rho_e{\bf E} = - \rho_e {\bf U \times B}/c$ is not negligible; however, it 
contributes to force balance across the field, but not to the
integrals (\ref{44}) and (\ref{45}).) 

For simplicity, we consider
just field lines near to the equator, with
  \eq \la{46}
  B_{\varpi} = \Phi/\varpi^2, \en
where $\Phi$ is a measure of the poloidal flux crossing the l-c.
The principal results are the prediction of the asymptotic values
\eq \la{47}
{U_{\varpi} \simeq c,\quad\quad U_{\phi}\simeq 0}, \quad\quad
B_{\phi} \simeq -(\alpha \varpi/c)B_{\varpi},\en
  and
  \eq \la{48}
\Gamma_ \simeq \left(\frac{\alpha^2 \Phi}{4\pi\eta
c^3}\right)^{1/3} \equiv \sigma^{1/3}. \en

The force-free approximation will remain valid as long as the
kinetic energy density is small compared with the electromagnetic,
i.e.
  \eq \la{49}
{\cal R} \equiv \frac{\Gamma nmc^2}{(B^2/4\pi)} \ll 1 \en
Substitution from (\ref{43}), (\ref{46}), (\ref{47}) and
(\ref{48}) yields
\eq \la{50}
{\cal R} \simeq \sigma ^{-2/3}
\simeq \Gamma^{-2}. \en
  The $\pm e$-density $n$ is written
conveniently as the product of a pair-production multiplicity
factor $M$ and an estimate for the local Goldreich-Julian density:
$ n = (M/e)(\rho_e)_{{\rm GJ}} \approx M (\alpha B_{p}/2\pi e c)$.
Substitution of numbers yields
\eq \la{51}
  \Gamma \simeq [(10^7/M)(B_{s12}R_{s6}^3/P_1^2)]^{1/3}\en
in terms of a stellar field $B_s = 10^{12}{\rm G}$, stellar radius
$R_s = 10^6$ and pulsar rotation period $P_1 = 1{\rm sec}$. Thus
even with $M= 10^3$, the predicted ${\cal R} \ll 1$ for the longest
periods, so that the basic approximation leading to (\ref{9}) and
therefore the solutions of Section 6 -- constructed subject to the
crucial equatorial boundary condition $B_z =0$ -- appears to
remain valid over the whole domain.

   The result is a self-consistency check for the present class of model,
but also shows up a basic limitation to its applicability to a
real pulsar; for the condition (\ref{49}) is -- not surprisingly
-- also the  condition that the energy flow at infinity be
primarily by the Poynting flux rather than the kinetic flux. In
fact, the wind flow results are highly model-dependent. Even with
the same equatorial boundary condition, the predicted dominantly
toroidal field at infinity may be dynamically unstable (e.g. Begelman 1998).
 Further, experience with other models with non-radial field lines beyond
the l-c (MRWW, FM, MS; Beskin et al 1993) suggests strongly that
in an alternative model with field lines approaching the equator
obliquely (cf Sub-section 7.1), particles streaming into the
equator will have acquired much higher $\Gamma$-values.

     Finally, we emphasize that this paper has been concerned with
the conditions for the construction of viable models. In practice,
the particle pressure requirements in the equatorial and dead
zones may very well put physical limitations on the model
parameters -- the extent $x_c$ of the dead zone and the associated
separatrix $P_c$.

{\bf ACKNOWLEDGEMENTS} The authors wish to thank the referee Professor 
D.B. Melrose for helpful suggestions on presentation.

      {\bf References}\\
Abramowitz, M., Stegun, I.A., 1965, (AS), Handbook of Mathematical
Functions. Dover, New York\\
Begelman, M.C., 1998, ApJ, {\bf 493}, 291\\
Beskin, V.S., Gurevich, A.V., Istomin, Ya. N., 1993, Physics of
the Pulsar Magnetosphere. Cambridge University Press\\
Biggs, J.D., McCulloch, P.M., Hamilton, P.A., Manchester, R.N.,
Lyne, A.G., 1985, MNRAS, {\bf 215}, 281\\
Blackman, E.G., Field, G.B. 1993, Phys. Rev. Lett., {\bf
71}, 3481\\
Contopoulos, I.,1995, ApJ, {\bf 446}, 67\\
Contopoulos, I., Kazanas, D., Fendt, C., 1999 (CKF), ApJ, {\bf 511},
351\\
Erd\'elyi A., Magnus W., Oberhettinger F., Tricomi F.G., 1954,
Tables of Integral Transforms. McGraw-Hill, New York\\
Fitzpatrick, R., Mestel, L., 1988a,b (FM), MNRAS, {\bf 232}, 277 and
303\\
Goldreich, P., Julian, W.H., 1969 (GJ), ApJ., {\bf 157}, 869 \\
Goldreich, P., Julian, W.H., 1970, ApJ, {\bf 160}, 971\\
Gradshteyn, I.S., Ryzhik, I.M., 1980, (GR), Table of Integrals,
Series and Products. Academic Press, New York\\
Holloway, N.J., Pryce, M.H.L., 1981, MNRAS, {\bf 194}, 95\\
Jeffreys, H., Jeffreys, B.S., 1972, Methods of Mathematical
Physics, 3rd Edition (Paperback). Cambridge University Press,
Cambridge\\
Li, J., Melrose, D.B., 1994, MNRAS, {\bf 270}, 687\\
Lynden-Bell, D., 1996, MNRAS, {\bf 279}, 389\\
Lyne, A.G., Manchester, R.N., 1988, MNRAS, {\bf 234}, 477\\
Melatos, A., Melrose, D.B., 1996, MNRAS, {\bf 279}, 1168\\
Mestel, L. 1999, Stellar Magnetism. Clarendon Press, Oxford, (Paperback
2003)\\
Mestel, L. 2001, Publications Astronomical Society
Australia, {\bf 18}, 1\\
Mestel, L., Pryce M.H.L., 1992 (MP), MNRAS, {\bf 254}, 355\\
Mestel, L., Shibata S., 1994 (MS), MNRAS, {\bf 271}, 621\\
Mestel, L., Spruit. H.C., 1987, MNRAS, {\bf 226}, 57\\
Mestel, L., Wang Y.-M., 1979 (MW), MNRAS, {\bf 188}, 799\\
Mestel, L., Robertson, J.A., Wang, Y.-M., Westfold, K.C., 1985
(MRWW), MNRAS, {\bf 217}, 443\\
Michel, F.C., 1969, ApJ, {\bf 158}, 727\\
Michel, F.C., 1973a, ApJ, {\bf 180}, L133\\
Michel, F.C., 1973b, ApJ, {\bf 180}, 207\\
Michel, F.C., 1974, ApJ, {\bf 192}, 713\\
Michel, F.C., 1991, Theory of Neutron Star Magnetospheres.
University of Chicago Press, Chicago\\
Rankin, J.M. 1993, ApJ Suppl, {\bf 85}, 145\\
Scharlemann, E.T., Wagoner, R.V., 1973, ApJ, {\bf 182}, 951\\
Shibata, S., Miyazaki, J., Takahara, F., 1998, MNRAS, {\bf 295},
L53; 2002, MNRAS, {\bf 336}, 233\\
Smith, I.A., Michel, F.C., Thacker, P.D., 2001, MNRAS, {\bf 322}, 209\\
van Leeuwen, A.G.J., Kouwenhoven, M.L.A., Ramachandran, R.,
Rankin, J.M., Stappers, B.W., 2002, Astr $\&$ Ap, {\bf 387}, 169\\
Vivekanand, M., Joshi, B.C., 1997, ApJ, {\bf 477}, 431\\
Wright, G.A.E., Fowler, L.A., 1981, in IAU Symposium 95 {\em
Pulsars}, eds W. Sieber, R. Wielebinski, Dordrecht: Reidel, 221\\

{\bf Appendix: Fourier methods}

{\bf The present models}

Other studies of the pulsar magnetosphere have written the flux
function $P$ as a Fourier integral in $z$. In the illustrative
MMWP model studied in MW and MP, in which the dead zone extends to
the l-c, $P$ is written as a Fourier cosine integral, appropriate
to a domain with the equatorial boundary condition $B_x \propto \p
P/\p z = 0$. In the present paper, with the perfectly conducting
wind domain extending from $x_c \le 1$ to $\infty$ and so with the
equatorial boundary condition $B_z \propto \p P/\p x =0$, the
appropriate form for the normalized $P$-function in the domain
$x_c \le x \le \infty$ is the Fourier sine integral
  \eq \la{A2}
  P = \frac{2}{\pi}\int_0^{\infty}f(x,k)\sin kz {\rm
d}k,\quad f=\int_0^{\infty} P\sin kz {\rm d}z. \en

 This is a Fourier
representation of the function $P(x,z)$ in the northern hemisphere
and $-P(x,z)$ in the southern, with $P(x, 0\pm) = \pm1$. By
Fourier's Theorem (e.g. Jeffreys and Jeffreys (1972)), at the
discontinuity on the equator, the Fourier representation
(\ref{A2}) should indeed have the value $[(+1) + (-1)]/2 = 0$.
However, this representation will exhibit the Gibbs phenomenon,
and may make difficult the accurate construction of behaviour near
the equator.

It is convenient to introduce also $F(x,k)$ defined by 
\eq \la{A3} 
f = \frac{1}{k} + F(x,k). \en
Because of the boundary
condition $P=1$ on $z=0$, the Fourier cosine transform of $\p P/\p
z$ is
\eq \la{A4}
\int_0^{\infty}\frac{\p P}{\p z} \cos kz {\rm
d}z = -1 + k\int_0^{\infty}P \sin kz {\rm d}z = - 1 + kf(x,k) =
kF(x,k). \en
Likewise, application of the Fourier sine transform to the
force-free equation (\ref{30.1}) yields
\eq \la{80}
  (1 - x^2)F^{\prime \prime} -
\frac{(1+ x^2)}{x}F^{\prime}  - (1 - x^2)k^2F = - g(x,k), \en
where as usual $F^{\prime} = F_x$, and
\eq \la{90}
g(x,k) =
\int_0^{\infty}S\frac{{\rm d}S}{{\rm d}P}\sin kz{\rm d}z. \en
In general, the integral (\ref{90}) can be converted into one over
$P$:
\eq \la{145.1}
  g(x,k) = \int_1^{0} \left(S\frac{{\rm
d}S}{{\rm d}P}\right)\frac{\sin kz}{(\p P/\p z)_x}{\rm d}P. \en

For a simple illustration, consider the Michel form (\ref{19}) for
$S(P)$, for which (\ref{30.1}) has the known solution (\ref{18})
(again normalized with $P_c = 1$), valid everywhere in the absence
of a dead zone (i.e. with $x_c =0$). We quote the known Fourier
transformation (AS, 9.6.25)
  \eq
\label{99}
  \int_{0}^{\infty}\frac{\cos kz}{(x^{2} + z^{2})^{\nu + 1/2}}{\rm d}z =
  \pi^{1/2}\left(\frac{k}{2x}\right)^{\nu}\frac{{\rm
K}_{\nu}(kx)}{\Gamma(\nu +1/2)}.
  \en
Then substitution from (\ref{20}) into (\ref{90}) and from
(\ref{18}) into (\ref{A2}) and (\ref{A3}) yields after integration
by parts
  \eq
  \la{64.5}
  F(x,k) = -x{\rm K}_1(kx),\quad \quad g(x,k) = 2kx^2{\rm K}_0(kx),
\en
  whence (\ref{80}) is seen to be satisfied from the known
properties of Bessel functions.

    Now return to the problem with a dipolar field on the star and
so with a finite dead zone, terminating at the point $(x_c, 0)$.
It emerges from the computations reported in Section 6 that even
with $x_c$ close to unity, the allowed $S(P)$ relations do not
differ much from the normalized Michel form
\eq \la{64.6}
  S = -2P + P^2.
\en 
The appearance of the neutral point $(x_c, 0)$ appears to be  associated 
rather  with the marked deviation from the radial structure for 
${\bf B}_{{\rm p}}$, 
found as the l-c is approached from without. It is of interest to discuss this
by application of the Fourier formalism.

 Having chosen a constructed model, and adopted the values found for
$P(1,z)$ and $P_x(1,z)$,
one then moves inwards from $x=1$, keeping $P$ and $P_x$
continuous, but using the $S(P)$ relation (\ref{64.6}). One can
write the solution of (\ref{80}) as 
\eq\la{65}
 F(x,k) = b_kF_{cf} +F_{pi}. \en 
 where $b_kF_{cf}$ is a complementary function, satisfying (\ref{80}) with zero
r-h-s, and $F_{pi}$ is a particular integral, conveniently constructed
to satisfy $F_{pi}(1,k) = 0$.
As in the MMWP problem, 
$F_{cf}$ is started off from $x=1$ by the
non-singular series
 \eq \la{105} F_{cf} = 1 + \frac{k^2}{4}(1-x)^2
+ \frac{k^2}{36}(1 - x)^3 + ...., \en 
and is then continued
inwards numerically. (Note that the function $F_{cf}$ for given
$k$ will remain unchanged during the whole operation.) 
 
With $g(x,k)$ prescribed and $F_{cf}$ known, the
particular integral is given by
\eq \la{107}
F_{pi} = F_{cf}\int_x^1{\rm
d}x\left[\frac{x}{(1+x)F_{cf}^2} \frac{(\int_1^x
[g(v)F_{cf}(v)/v]{\rm d}v)}{(1-x)}\right]. \en
Near the l-c, $F_{pi}$ is best computed from the series
\eq \la{110}
F_{pi} = -\frac{g(1,k)}{2}(1 - x) +
\frac{g^{\prime}(1,k)}{8}(1 - x)^2 + ......
  \en
found simply from (\ref{80}) and its first derivative.

With the particular choice for the P.I.,
  $F(1,k) = b_k F_{cf}(1,k) + F_{pi}(1,k) = b_k$,
so $b_k$ is fixed for given $S(P)$ from the previously constructed
solution for $x > 1$. For the present problem, from the given
values of $P$ on the l-c, continuity of $F_x$ and so of $P_x$ is
ensured by the construction of $F_{pi}$. Numerical continuation of
$F_{pi}$ from (\ref{107}) will depend on substitution into $g$ of
the values for $P$ emerging from the inward integration.

  The model with $x_c = \cdot 86$ was chosen for study. The inward 
integration is to be halted when a value $x = x_c$ is
reached for which, using (\ref{A4}) together with (\ref{65}),
\begin{eqnarray}
\la{130}
  0 &=& (\pi x_c/2)B_x(x_c, \,0)= (\pi/2)\left[\frac{\p P}{\p
z}\right]_{z=0} =
\int_0^{\infty}kF(x_c,\,k){\rm d}k  \\
&=& \int_0^{\infty}\left[kb_kF_{cf}(x_c,k) +
kF_{pi}(x_c,k)\right]{\rm
d}k \nonumber \\
\la{140} &=& \int_0^{\infty}k\left[F(1,k)F_{cf}(x_c,k) +
F_{pi}(x_c, k)\right]{\rm d}k.
\end{eqnarray}
   However, the numerical work yielded only partially satisfactory results. 
The preliminary integrations confirmed that the complementary function 
dominates -- i.e., the solution is indeed determined primarily by the 
distribution of $P$ on the l-c -- but
the predicted value for $x_c$ was $\approx .82$. Attempts to bring $x_c$ 
closer to the value
.86 found in the global integration of Section 6 by including higher 
$k$-values led to 
oscillating or unstable back transforms. We conclude that in a problem with 
the boundary condition $P_x(x,0) = 0$, requiring the Fourier sine 
integral (\ref{A2}), but with $P(x,0)=1$, the method is in general 
inappropriate for study of the field structure near the equator. 
 
 {\bf The equatorial boundary condition $B_x = 0$}

  For the models of this paper, recall that in the inner domain, the dead zone
terminates within the l-c at the point N with coordinates
$(x_c,\,0)$ with $x_c \le 1$. Within the dead zone the field lines
cross the equator normally, so that $\p P/\p z=0$ for $x < x_c$,
$z=0$; whereas the condition that the wind zone be perfectly
conducting enforces a fully open field crossing the l-c, so that
$\p P/\p x=0$ for $x > x_c$, $z=0$. This in turn yields a non-zero
$S$ both at the outer domain equator and on its continuation as
the separatrix $P_c$ between the wind and dead zones, so that much
of the current returns to the star as a sheet. Simultaneously, the
conditions of equilibrium require a gas pressure both at the
equator beyond $x_c$ and within the dead zone.

In the somewhat more complicated model of MS there is again a
force-free domain extending beyond the l-c, but now the boundary
condition $\p P/\p z \propto B_{x}(x,0) = 0$ is supposed to hold
in the outer domain $x>1$ also. This is the appropriate
approximation if e.g. a strong dynamically-driven macro-resistivity,
allows the radial wind flow along the equator to cross the field
lines. As in MW and MP, the solution for $P$ can now be written as
a Fourier cosine integral. The constraint (\ref{6}) still holds
away from the equator, but a local departure from torque-free
conditions, associated with the flow of gas within the equatorial
zone, allows $S(P)$ to vary along the equator. Further, with
$B_x(x,0) = 0$, the arguments in Section 3 requiring a non-zero
$S$ at the equator and so along the separatrix between the wind
and dead zones no longer hold. It is tentatively postulated that
the current closure condition
can now be satisfied by the simple choice $S(P_{c})=0$, allowing
$S(P)$ to go to zero continuously, without there needing to be a
sheet current along $P_c$. There is no pinched, high pressure
equatorial sheet, nor is there a mandatory introduction of
pressure into the dead zone.

   However, as noted in Section 7, these simplifications come at a
price: for
the case studied, with no current sheet on the separatrix, and
with the equatorial boundary condition $\p P/\p z \propto B_x=0$
holding everywhere, a solution that is well-behaved and continuous
at the l-c blows up before it can reach infinity. In MP, it is
shown that imposition of smooth continuity at the l-c yields a
singularity in the solution at less than two l-c radii.
Equivalently, one can apply standard asymptotic theory to the
Fourier integral, demanding that the Fourier transform behave
properly at infinity; there then appears an incompatibility of
sign between solutions within and without the l-c (Mestel 2001).
Resolution of the dilemma may be as suggested in MP, with local
breakdown in the simple plasma condition ${\bf E}\cdot{\bf B} =0$, not
only
on the equator, but also in a thin, dissipative volume domain,
idealized as a cylindrical shell symmetric about the rotation
axis. Further work is needed to test whether instead, relaxation
of the condition $S(P_c) = 0$ will allow construction of a global
solution of (\ref{30.1}) subject to the condition $B_x(x,0) = 0$.

\end{document}